\documentclass[12pt,twoside]{article}
\usepackage{fleqn,espcrc1}
\usepackage{epsfig}

\newcommand{\AmS}{{\protect\the\textfont2
  A\kern-.1667em\lower.5ex\hbox{M}\kern-.125emS}}

\title{Limits Due to Instrumental Polarisation in CMB Experiments at Microwave
       Wavelengths}
\author{E.~Carretti\address[TESRE]{Istituto Te.S.R.E.--C.N.R., 
             Via Gobetti 101, I-40129 Bologna, ITALY},
	R. Tascone\address[IRITI]{I.R.I.T.I.--C.N.R., 
	     C.so Duca degli Abruzzi 24, I-10129, Torino, ITALY},
	S. Cortiglioni\addressmark[TESRE]
        J. Monari\address[IRA]{I.R.A.--C.N.R., Via Gobetti 101, 
	I-40129 Bologna, ITALY}
        and
        M. Orsini\addressmark[TESRE]\address[DipAstr]{Dipartimento di 
	     Astronomia--Universit\`a di Bologna, 
	     Via Ranzani 1, I-40127 Bologna, ITALY}}

\begin{document}

\maketitle

\begin{abstract}
An extended analysis of some instrumental polarisation sources has 
been done, as a consequence of the renewed interest in extremely sensitive polarisation 
measurements stimulated by Cosmic Microwave Background experiments. 
The case of correlation polarimeters, being them more suitable than other
configurations, has been studied in 
detail and the algorithm has been derived to calculate their intrinsic 
sensitivity limit due to device characteristics as well as to 
the operating environment. The atmosphere emission, 
even though totally unpolarised, has been recognized to be the 
most important source of sensitivity 
degradation for ground based experiments.
This happens through receiver component 
losses (mainly in the OMT), which generate instrumental polarisation in genuinely uncorrelated
signals. The relevant result is that, also in best conditions (cfr. Antarctica),
integration times longer than $\sim 40$~s are not allowed on ground without
modulation techniques. Finally, basic rules to estimate the maximum modulation
period for each instrumental configuration have been provided. 
\end{abstract}

\section{Introduction}\label{intro}
The Cosmic Microwave Background (CMB), which has a
black--body (BB) spectrum at a thermodynamic temperature 
$T_{\rm o}=2.728$K (Fixsen et al., 1996), is a powerful tool to
understand origin and evolution of our universe. The CMB looks almost 
isotropic and unpolarised
and any detection of deviations from an ideal BB would be very
important because they allow the estimate of cosmological parameters
(Jungman et al. 1996, Zaldarriaga et al. 1997, Efstathiou \& Bond
1999). Very small temperature fluctuations in the CMB have been detected
at both large (Smooth et al. 1991, Bennet et al. 1996) and small (De
Bernardis et al. 2000, Hanany et al. 2000, Miller et al. 1999)
angular scales, but only upper limits on the CMB polarisation (CMBP)
have been evaluated 
(Hedman et al. 2000 and see Staggs et al. 1999 for a complete list of the
CMB polarisation measurements).

The analysis of the linearly polarised component
of the CMB promises to be very important
in the cosmological frame. In fact, the CMBP
can solve the degeneracies
among cosmological parameters that CMB anisotropy alone is not able to remove
(Zaldarriaga et al. 1997). Moreover, it allows the separation of
scalar and tensorial
components of the primordial fluctuations providing a way to disentangle among
different inflationary models (Kamionkowski \& Kosowsky 1998).

Several microwave experiments, based on radiometric receivers,
are either operating or planned 
to search for the CMB linear polarisation
at both large and small angular scales, either from ground or from space.
SPOrt{\footnote{see SPOrt home page 
                http://sport.tesre.bo.cnr.it}} (Cortiglioni et al. 1999), 
the Milano Polarimeter
(Sironi et al. 1998) and 
POLAR{\footnote{see POLAR home page 
                http://cmb.physics.wisc.edu/polar/}}
(Keating et al. 1998) will
operate at $7^{\rm o}$ scale by using simple optics;
PIQUE{\footnote{see PIQUE home page 
                http://dicke.princeton.edu/~pique/pique.html}} 
(Hedman et al., 2000) will investigate angular scales below $1^{\rm o}$
by means of mirror optics.
Both MAP{\footnote{see MAP home page 
http://map.gsfc.nasa.gov/}} (Bennet et al. 1997, Wright 1999)
and Planck--LFI{\footnote
{see Planck home page http://astro.estec.esa.nl/SA-general/Projects/Planck/}}
(Tauber, 2000), even though they do not have polarimeters on board, will also
try to derive CMBP information at subdegree angular scale.
For all of them the expected signal level is low:
on $7^{\rm o}$ scale the polarised
emission can be lower than 1~$\mu$K,
while on smaller angular scales ($< 1^{\rm o}$) signals can be of the order of
few $\mu$K.

Moreover, it is well known that CMB experiments have to deal with Galactic
contribution removal. In fact, the Galactic background 
beside its intrinsic interest, acts as a foreground
for CMB experiments and only its accurate knowledge will allow detections
of the CMB features (Brandt et al. 1994, Dodelson 1997, Bouchet et al.
1999, Tegmark et al 2000, Prunet et al. 1998, Kogut \& Hinshaw 2000, Tucci et
al. 2000a, Baccigalupi et al. 2000).
Our Galaxy is featured by a smoothed linearly polarised
background emission,
carrying information on the Galactic structure, that has been observed
at frequencies up to 2.7 GHz (Brouw \& Spoelstra 1976, Junkes et
al. 1987, Duncan et al. 1999, Duncan et al. 1997, Uyaniker et al. 1999,
McClure-Griffiths et al. 2001, Gaensler et al. 2001), where
it results to be dominated by the synchrotron emission. 

\begin{figure}
\includegraphics[scale=0.39]{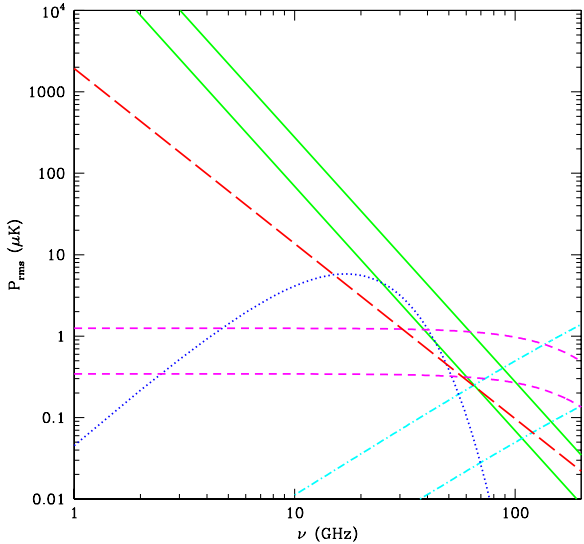}
  \caption{Plot of the expected level of polarisation emission on 
  $7^{\rm o}$ angular scale: CMB (dashed lines:
  two CDM models with optical depth $\tau$=0.1--0.5,
  respectively), Synchrotron (solid lines: normalized to 5--20~mK at 2.4~GHz),
  Free--Free (long--dashed), Thermal Dust (dot--dashed lines: 5--0.5\% polarised)
  and Spinning Dust (dotted). See Tucci et
  al. (2000b) for details about normalizations.    
  }
\label{foreFig}
\end{figure}

No all--sky observations of polarised
Galactic background have been made so far in the microwave domain,
where synchrotron emission should prevails,
even though  it has been recently 
suggested that spinning and magnetic dust
grain linearly polarised emission should be 
not negligible in the 20--70~GHz range
(Draine \& Lazarian 1999, Lazarian \& Draine 2000, Tegmark et al. 2000).
At millimeter wavelengths, saying above
100~GHz (Kogut et al. 1996), the Galactic emission is dominated by
Galactic thermal dust.
The expected scenario, showing that in any case the antenna temperatures to be
detected are extremely low, is represented in Figure~\ref{foreFig}.

In spite of this, most of the noise in CMBP experiments would come from the
CMB itself, because such experiment  must detect a
very weak signal (few $\mu$K or less) in presence of
the strong unpolarised background $T_{\rm o}$. In addition,
the instrumental noise
and the atmospheric emission (for ground based experiments) contribute to the
total signal collected by the instrument making stronger and variable this
unpolarised (unwanted) component.

The losses of radiometer components will transform such an 
unpolarised part of the input signal in a so called
{\it spurious} polarised component, 
which should be kept, in principle, lower than the
{\it genuine} polarised signal to be searched (namely
lower than 1~$\mu$K). Every
experiment has to deal with such instrumental offsets, which are also
detected as {\it genuine} polarisation, but they are not.
Then, even though in certain conditions
suitable scanning strategies  or
signal modulation techniques may allow the extraction of the true signal
by post--processing data analysis (for
instance cfr. Revenu et al. 2000, Carretti et al. 2000, Carretti et al. 2001),
the common baseline is to keep offsets as low as possible.

However, the most important problem introduced by the offset are its
fluctuations, which can strongly degrade the instrument sensitivity.
The radiometer sensitivity equation may be written as
(Wollack \& Pospieszalski 1998, Wollack 1995):
\begin{equation}
  \Delta T_{\rm rms} = \sqrt{
                      {k^2 \,T_{\rm sys}^2 \over \Delta\nu\,\tau}+
               T_{\rm offset}^2
               \left({\Delta G \over G}\right)^2 +
               \Delta T_{\rm offset}^2
            }
\label{trmseq}
\end{equation}
being $T_{\rm sys}$, $T_{\rm offset}$ and $\Delta T_{\rm offset}$ the system
temperature, the offset equivalent temperature and its fluctuation,
respectively; $G$ is the radiometer
gain, $\tau$ the integration time, $\Delta \nu$ the radiofrequency bandwidth
and $k$ a constant depending on the
radiometer type.
The first term represents the white noise of an ideal and stable
radiometer;
both the second and the third terms
represent additional noise generated by instrument instabilities
driven by the offset.

The main aim of this work is to analyse some usual configuration of microwave
polarimeters with respect to the offset
generation, to evaluate the effect on the sensitivity
degradation introduced by the offset itself. This analysis
considers only correlation polarimeters, which are intrinsically more stable
with respect to other configurations.

\section{${\bf Q}$ and ${\bf U}$ Stokes Parameters Measurements}\label{twopol}

The polarisation status of the radiation is usually described by the Stokes
parameters $I$, $Q$, $U$ and $V$ (see, for instance, Kraus 1987, Rohlfs \&
Wilson 2000).
The linearly polarised component is fully defined
by $Q$ and $U$, which give the linear polarised intensity and
the polarisation angle:
\begin{eqnarray}
      I_p &=& \sqrt{Q^2 + U^2} \\
   \alpha &=& {1\over 2}\tan^{-1}(U/Q).
\end{eqnarray}
$V$ and $I$ represent the circularly polarised component and
 the total emission (unpolarised~+~polarised), respectively.

Since both the CMB and the Galactic emission are
expected to be (partially) linearly polarised, with negligible circular
polarisation ($V\approx 0$), this paper
deals only with the $Q$ and $U$ detection.

$Q$ and $U$ Stokes parameter measurements using correlation
instruments can be carried on in several ways (for instance see
SPOrt, Milano Polarimeter, POLAR, PIQUE). However, all of them
multiply the two signals, saying $A$ and $B$ coming from
the antenna system, providing two outputs
that, using a Fourier spectral representation, can be expressed
as:
\begin{eqnarray}
  O_1 &\propto& 2\, \Re (A B^*) \nonumber\\
  O_2 &\propto& 2\, \Im (A B^*)
  \label{o1o2eq}
\end{eqnarray}
These instruments can be divided into two groups
with respect to the signals to be correlated,
depending on the characteristics of the antenna system. 
In fact, the two signals $A$ and $B$ can represent either the two
circularly or the two
linearly polarised components of the incoming radiation.

In the case where $A$ and $B$ are the right--hand and
left--hand circularly polarised components, respectively, the two
correlator outputs are proportional to $Q$ and $U$:
\begin{eqnarray}
  O_1 &\propto& 2\, \Re (E_R E_L^*) = Q\nonumber\\
  O_2 &\propto& 2\, \Im (E_R E_L^*) = U
  \label{o1o2lreq}
\end{eqnarray}
and they provide a direct and a simultaneous measurement of the two Stokes
parameters describing the linear polarisation.

Differently, if $A \propto E_X$ and $B \propto E_Y$ are the two
linearly polarised components with respect to the cartesian reference frame
of the instrument, the two correlator outputs are proportional to
$U$ and $V$:
\begin{eqnarray}
  O_1 &\propto& 2\, \Re (E_X E_Y^*) = U\nonumber\\
  O_2 &\propto& 2\, \Im (E_X E_Y^*) = V
  \label{o1o2xyeq}
\end{eqnarray}
providing only one (U) of the two Stokes parameters needed to describe
the linear polarisation.
However, a rotation
of the reference frame converts $Q$ into $U$ and viceversa through:
\begin{eqnarray}
  Q &=& Q_{0} \cos(2\beta) + U_{0} \sin(2\beta), \\
  U &=& U_{0} \cos(2\beta) - Q_{0} \sin(2\beta),
  \label{quroteq}
\end{eqnarray}
where $\beta$ is the rotation angle of the antenna around its
axis. In this way, the measurement of $Q$ can be done with
$\beta = 45^{\rm o}$, so that the two
parameters can be obtained sequentially. Since this configuration  does not
provide simultaneously $Q$ and $U$ its time
efficiency is reduced of 50\% with respect to the previous
configuration, where  circularly polarized components are used.

\section{Spurious Polarisation}\label{spu}

In the previous section the outputs for an ideal polarimeter
have been described.
In real instruments the two channels $A$ and $B$ are not completed isolated.
Hence, they are partially
correlated also in presence of unpolarised radiation only.

\begin{figure}
\includegraphics[angle=0,scale=1.0]{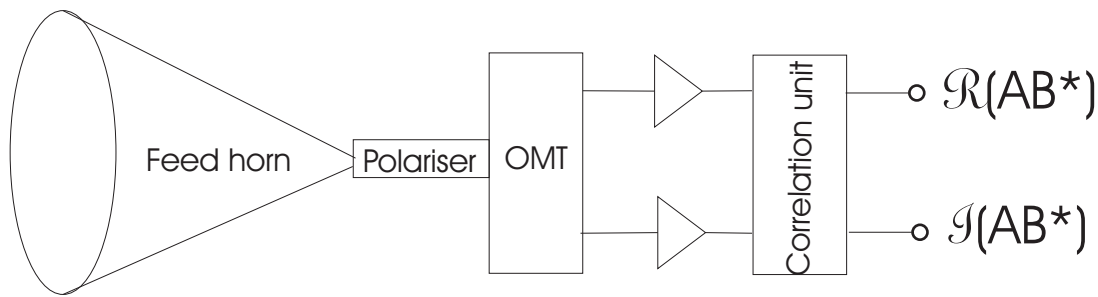}
  \caption{A correlation polarimeter
  can be divided in three parts. The antenna
   system collects the sky signal and provides the signals $A$ and $B$
   (the polariser is requested only to have circularly polarised components).
   The amplification chain amplifies the two divided signals. Finally, the
   correlation unit multiply the signals and provides two outputs proportional
   to $\Re(AB^*)$ and $\Im(AB^*)$. Antenna system and correlation unit
   represent the common parts where $A$ and $B$ can cross--correlate.
  }
\label{fig1}
\end{figure}

Such a contamination can be generated only in common parts, that is where
signals propagate together (in the same device) and they can cross each
to the other.
In correlation polarimeters the common parts are 
the antenna system as well as the correlation unit (see Figure~\ref{fig1}).

In order to estimate how these effects can influence the measured
outputs $Q'$ and $U'$, they will be analyzed with respect to the
actual (from the sky) values of the whole Stokes parameters set ($I$, $Q$,
$U$, $V$). 

The contaminated signals $A'$ and $B'$ at the output of a device
can be expressed, with respect to the ideal ones $A$ and $B$, as
\begin{eqnarray}
  A' &=& A + b\, B,\\
  B' &=& a\, A + B,
  \label{ABconteq}
\end{eqnarray}
where $a$ ($b$) represents the fractional contamination of $A$ ($B$) in $B'$
($A'$).
Since spurious
contributions can be also phase shifted with respect to the parent signals, 
both $a$
and $b$ are complex values.

By substituting $A'$ and $B'$
in equations~\ref{o1o2eq}, through equations~\ref{o1o2lreq} (circularly
polarised components), the measured
$Q'$ and $U'$ after correlation are
\begin{eqnarray}
  Q' &=& Q\left[ 1 + \Re(a\,b\,^*)/2 \right] \nonumber\\
     &-& U\left[ \Im(a\,b\,^*)/2 \right] \nonumber\\
     &+& I\,\,\left[ \Re(a\,) + \Re(b\,)\right] \nonumber\\
     &+& V\left[ \Re(a\,) - \Re(b\,)\right],\\
     & &      \nonumber\\
  U' &=& U\left[ 1 - \Re(a\,b\,^*)/2 \right] \nonumber\\
     &-& Q\left[ \Im(a\,b\,^*)/2 \right] \nonumber\\
     &-& I\,\,\left[ \Im(a\,) - \Im(b\,)\right] \nonumber\\
     &-& V\left[ \Im(a) + \Im(b)\right],
  \label{QUcConteq}
\end{eqnarray}
That clearly shows
$Q'$ and $U'$ be contamined by the all Stokes parameters. Since
$|a|, |b| << 1$ and $V \sim 0$ (no circular polarisation), the most relevant
contamination is due to $I$. This is particularly true
for CMBP experiments, where $I$ (i.e. the total emission)
is $6\div 7$ order of magnitude
greater than (expected) $Q$ and $U$.

In the case of linearly
polarised components (equations~\ref{o1o2xyeq}), the quantity $U'$ is given by
\begin{eqnarray}
  U' &=& U\left[ 1 + \Re(ab^*)/2 \right] \nonumber\\
     &-& V\left[ \Im(ab^*)/2\right] \nonumber\\
     &+& I\,\,\left[ \Re{(a)} + \Re{(b)}\right] \nonumber\\
     &+& Q\left[ \Re{(a)} - \Re{(b)}\right],
  \label{Ulconteq}
\end{eqnarray}
Even though both
$I$ and $Q$ have first order coefficients, $I \gg Q$, and also in this
case the most relevant contamination comes from $I$.
Hence, hereinafter,
 {\it spurious polarisation} will denote
the offset term due to $I$.

However, it is possible to reduce the spurious polarisation effect on
the overall instrument performances, for example
by using modulation techniques. Internal modulations 
are quite common and they can reduce the problem of spurious polarisation
inside the correlation unit. External modulations are more
difficult to realise and may also introduce some problems, such as from 
by either spillover changing or chopping reflectors (Cortiglioni
1995). For these reasons the antenna system will be carefully analysed 
in the following sections.

\section{Spurious Polarisation from the Antenna System}\label{spAnSy}

 Hereinafter,
 the antenna system will include: the feed horn, the polariser
(when circularly polarised components are used) and the OMT (Orthomode
Transducer), which separates the two components to be correlated.
In order to highlight the effects of the $I$ parameter,
the sky signal entering the antenna will be considered to be unpolarised 
and also perfectly matched devices are assumed. In the following, 
the correlation of both circularly and linearly polarised components 
are considered.

\subsection{Correlation of Circularly Polarised Components}

Usually, the two circular components are
obtanined  by means of a waveguide  polariser and an  OMT (see, for
instance,  SPOrt and the Milano Polarimeter):  the two linear
components $E_X$ and $E_Y$ are collected by the feed horn and
the polariser inserts a $90^{\rm o}$ phase delay between them. Hereafter
the Fourier spectral representation will be used and 
the polariser outputs can be expressed as
\begin{eqnarray}
  E_{\parallel}    &=& E_X, \nonumber \\
  E_{\perp}    &=& j \; E_Y,
  \label{apar_aort_eq}
\end{eqnarray}
where the common phase delay has been omitted.
The OMT separates the two orthogonal components.
However, since the OMT reference frame is $45^{\rm o}$ rotated 
with respect to the polariser, the OMT outputs are
\begin{eqnarray}
  A    &=& {1\over \sqrt{2}} (E_{X} - j\,E_{Y}) = E_L, \nonumber\\
  B    &=& {1\over \sqrt{2}} (E_{X} + j\,E_{Y}) = E_R,
  \label{abLReq}
\end{eqnarray}
which are  the left--hand and the right--hand circularly polarised
components. After the  amplification,  the OMT outputs are
multiplied by
the correlation unit as in equations~\ref{o1o2eq}
and~\ref{o1o2lreq} (see Figure~\ref{fig1}). 
In a real instrument there are
deviations from this ideal  situation, originating in each single 
device (feed horn, polariser and OMT), as it will be discussed 
below.

\subsubsection{The Feed Horn}

\begin{figure}
\includegraphics[scale=0.6]{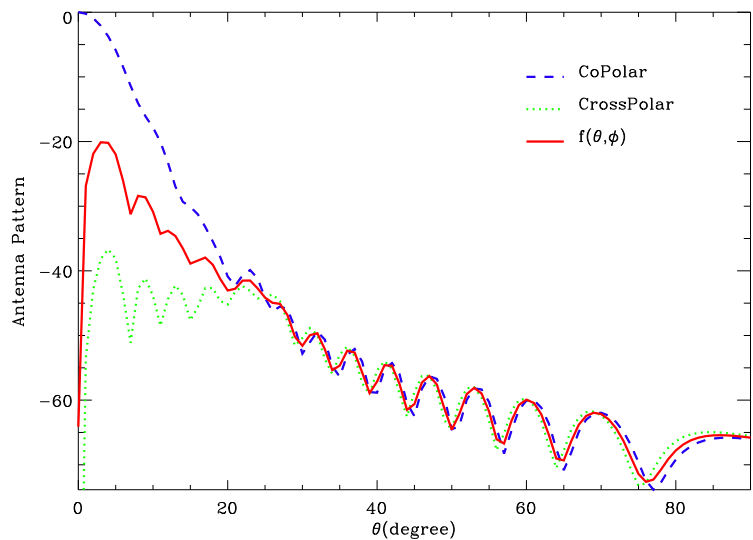}
  \caption{Co--polar, Cross--polar and function $f$ (see text) patterns for the
  feed horn of the SPOrt experiment at 22~GHz in the $45^{\rm o}$ direction.
  }
\label{figPatt}
\end{figure}

\begin{figure}
\includegraphics[scale=0.6]{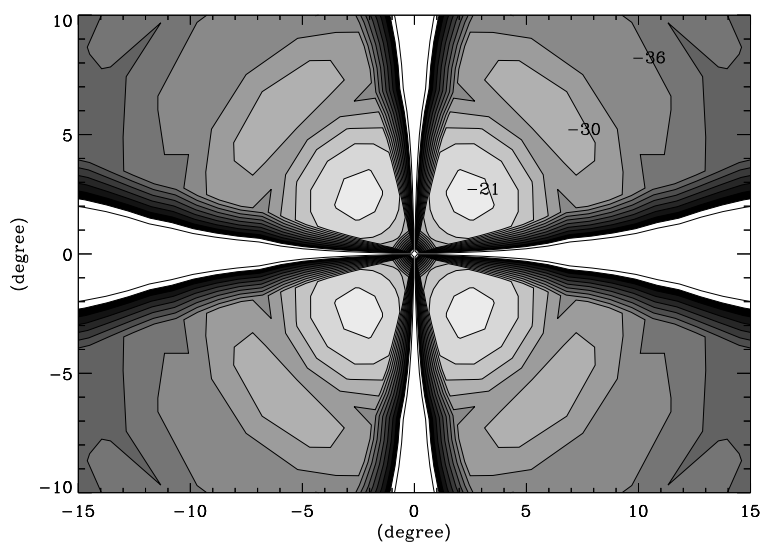} 
  \caption{Contour plot of the function $f$ for the 22~GHz 
  feed horn of SPOrt (HPBW~$=7^{\rm o}$).
   Levels are in dB (3~dB by 3~dB) and the maximum level is -21~dB (the 4 main
   lobes). The function is dominated
  by a structure with 4 lobes, whose sizes are $\sim 5^{\rm o}$ and 
  whose distances are $\sim 7^{\rm o}$. 
  As a consequence, the anisotropy smoothed on the
  $\sim$HPBW angular scale is responsible for the
  instrumental polarisation by the feed horn.
  }
\label{figCrossCont}
\end{figure}

The unpolarized radiation 
(sky and atmospheric emission) is described by the spectral
distribution of the incident electric field
$\underline{E}(\theta,\phi)$, 
where $\theta$ and $\phi$ are
the spherical angular coordinates.
The signals at the feed horn output can be expressed as
\begin{eqnarray}
V^F_{\parallel} &=& \int_0^{2\pi}\int_0^\pi
                    \underline{E}(\theta,\phi) \cdot 
                    \underline{h}_{\parallel}(\theta,\phi) \;
                            \sin\theta\,d\theta\,d\phi + N^F_{\parallel}\nonumber\\
V^F_{\perp}     &=& \int_0^{2\pi}\int_0^\pi
                    \underline{E}(\theta,\phi) \cdot 
                    \underline{h}_{\perp}(\theta,\phi) \;
                            \sin\theta\,d\theta\,d\phi + N^F_{\perp},
\end{eqnarray}
where the symbols $\parallel$ and $\perp$ denote the principal
directions of the polariser, $N^F_{\parallel}$ and $N^F_{\perp}$ 
are the (thermal) noise generated by the feed horn and the vector quantities 
$\underline{h}_{\parallel}$ and
$\underline{h}_{\perp}$ are the effective heights of the antenna  for
the $x$ and $y$ polarizations, respectively, and corresponding to
directions $\theta$, $\phi$.
However, the antenna pattern can introduce some correlation, giving rise to
spurious polarisation.  This can be estimated by computing
\begin{eqnarray}
V^F_{\perp} {V^F_{\parallel}}^* &=& \int_0^{2\pi}\int_0^\pi
                           (\underline{E}\cdot\underline{h}_{\perp})
                           (\underline{E}\cdot\underline{h}_{\parallel})^*\;
                           \sin\theta\,d\theta\,d\phi 
			   \nonumber\\
	     &=& \int_0^{2\pi}\int_0^\pi
	         \left[|E_{\theta}|^2 h_{\perp\theta}h_{\parallel\theta}^* +
	               |E_{\phi}|^2 h_{\perp\phi}h_{\parallel\phi}^*
		 \right]\;
                 \sin\theta\,d\theta\,d\phi 
\end{eqnarray}
where cross--products are null due to the uncorrelation among
$E_{\theta}$, $E_{\phi}$, $N^F_{\parallel}$ and $N^F_{\perp}$.
In terms of equivalent antenna temperature the correlated instrumental
noise can
be expressed as (see Appendix~\ref{app1} for details)
\begin{eqnarray}
V^F_{\perp} {V^F_{\parallel}}^* &=&\int_0^{\pi}\sin\theta\,d\theta
                                    \int_0^{\pi/2}d\phi\,
                                   \left[\Delta T_b(\theta,\phi)-
				         \Delta T_b(\theta,\phi+\pi/2)+
				   \right.\nonumber\\
				& & \;\;\;\;\;\;\;\;\;\;\;\;\;\;\;\;
				    \;\;\;\;\;\;\;\;\;\;\;\;\;\;\,
				    \left. \Delta T_b(\theta,\phi+\pi)-
				           \Delta T_b(\theta,\phi+3/2\pi)
                                                 \right] f(\theta,\phi)
\label{crossFeedeq}
\end{eqnarray}
where
\begin{equation}
f(\theta,\phi) = -F_{\rm e}(\theta, \phi)F_{\rm o}^*(\theta, \phi+\pi/2) +
                  F_{\rm o}(\theta, \phi)F_{\rm e}^*(\theta, \phi+\pi/2)
\label{crossFeedeq1}
\end{equation}
and where $\Delta T_b(\theta, \phi)$ is the brightness temperature 
anisotropy distribution 
of the sky and $F_{\rm e}$ and $F_{\rm o}$ are the co--polar and cross--polar
radiation patterns for a linearly polarised feed horn, respectively. 
The relevant result is that the
correlation term depends on the radiation anisotropy rather than on the
total intensity and that the term $f(\theta,\phi)$
represents the non ideality of the feed horn from the instrumental correlation
points of view rather than the cross-polar pattern (see Figure~\ref{figPatt}).
Moreover, Figure~\ref{figCrossCont} shows that
$f(\theta,\phi)$ mainly consists of 4 lobes, whose size and distance are
roughly equal to the HPBW. That means the source of instrumental correlation is
the anisotropy smoothed on the HPBW angular scale.

By considering that the anisotropy of the CMB
is about 30~$\mu$K on $7^{\rm o}$ angular scale, 
the contribution of the feed horn can be considered negligible. In fact,
in the case of the example of the Figure~\ref{figCrossCont},
the integral value of the function $f(\theta,\,\phi)$ with respect to
the integral of the co--polar pattern $F_e(\theta,\,\phi)$ is
$\sim -25$~dB, which yields to a correlation term of the order of 
0.1~$\mu$K.

\subsubsection{Polariser}

The polariser can be
described by a 4 ports device whose outputs can be expressed in the reference
frame defined by its two principal linear polarisations as follows
\begin{eqnarray}
E_{\parallel} &=& S_{\parallel} \left( V^F_{\parallel} +
                                       N^{\rm pol}_{\parallel}
                   \right)  \nonumber\\
E_{\perp}     &=& S_{\perp} \left( V^F_{\perp} +
                                   N^{\rm pol}_{\perp} \right)
\end{eqnarray}
where $N^{\rm pol}_{\parallel}$ and $N^{\rm pol}_{\perp}$ are the noise
generated by the polariser and evaluated at its input. 
The parameters $S_{\parallel}$ and $S_{\perp}$ are the related transmission 
parameters.
In ideal conditions $S_{\parallel} = 1$ and $S_{\perp} = j$ (omitting
the common phase delay).

\subsubsection{OMT}
The OMT can be described as a 4 port  network as well. In the OMT
reference frame the input signals are:
\begin{eqnarray}
E_1 &=& \frac{1}{\sqrt{2}}  \left[ S_{\parallel} \left( V^F_{\parallel} +
                                       N^{\rm pol}_{\parallel}\right) -
                               S_{\perp} \left( V^F_{\perp} +
                                   N^{\rm pol}_{\perp} \right) \right] \nonumber\\
E_2 &=& \frac{1}{\sqrt{2}}  \left[ S_{\parallel} \left( V^F_{\parallel} +
                                       N^{\rm pol}_{\parallel}\right) +
                               S_{\perp} \left( V^F_{\perp} +
                                   N^{\rm pol}_{\perp} \right) \right]
\label{pomt}
\end{eqnarray}
and the output signals can be written as a linear combination of the
input ones:
\begin{eqnarray}
A &=& S_{A1}  (E_1 + N^{\rm OMT}_1) + S_{A2}  (E_2 + N^{\rm OMT}_2) \nonumber\\
B &=& S_{B1}  (E_1 + N^{\rm OMT}_1) + S_{B2}  (E_2 + N^{\rm OMT}_2)
\label{aomt}
\end{eqnarray}
where $S_{A1}$, $S_{A2}$, $S_{B1}$, $S_{B2}$ are the OMT transmission
parameters, that
in ideal conditions must be $S_{A1} / S_{B2} = 1$, $|S_{A1}| = 1$, 
and $S_{A2} = S_{B1} = 0$. $N_1^{\rm OMT}$ and $N_2^{\rm OMT}$ denote the
noise generated in the two OMT arms and carried to the OMT input.

From equations~\ref{pomt} and~\ref{aomt}, $A$ and $B$ become
\begin{eqnarray}
A &=& \frac{1}{\sqrt{2}}  \left[\right.
   \left( S_{A1} + S_{A2} \right)  S_{\parallel}  \left( N^{\rm pol}_{\parallel} +
    V^F_{\parallel} \right) -
   \left( S_{A1} - S_{A2} \right)  S_{\perp}  \left( N^{\rm pol}_{\perp} +
   V^F_{\perp} \right)
                        \left.\right]\nonumber\\
   & & + S_{A1}N_1^{\rm OMT} + S_{A2}N_2^{\rm OMT} \nonumber \\
\nonumber \\
B &=& \frac{1}{\sqrt{2}}  \left[\right.
   \left( S_{B1} + S_{B2} \right)  S_{\parallel}  \left( N^{\rm pol}_{\parallel} + V^F_{\parallel} \right) -
   \left( S_{B1} - S_{B2} \right)  S_{\perp}  \left( N^{\rm pol}_{\perp} +
    V^F_{\perp} \right)
                        \left.\right]\nonumber\\
  & & + S_{B1}N_1^{\rm OMT} + S_{B2}N_2^{\rm OMT}
\end{eqnarray}

\subsubsection{Contamination Terms}

The spurious polarization can be estimated by
computing  the product $AB^*$, which is the spectral density of the
spurious contribution to the measured polarization:
\begin{eqnarray}
A  B^*  & = & \frac{1}{2}
   \left( S_{A1}  S_{B1}^* + S_{A2}  S_{B2}^* \right)
    \left[
      \left| S_{\parallel} \right|^2
       \left(
         \left| N_{\parallel}^{\rm pol} \right|^2
         + \left| V_{\parallel}^{F} \right|^2
      \right)
       + \left| S_{\perp} \right|^2
       \left(
         \left| N_{\perp}^{\rm pol} \right|^2
         + \left| V_{\perp}^{F} \right|^2
      \right)
   \right] \nonumber \\
  & + &  \frac{1}{2} \left( S_{A1}  S_{B2}^* + S_{A2}  S_{B1}^* \right)
    \left[
      \left| S_{\parallel} \right|^2
       \left(
         \left| N_{\parallel}^{\rm pol} \right|^2
         + \left| V_{\parallel}^{F} \right|^2
      \right)
      - \left| S_{\perp} \right|^2
       \left(
         \left| N_{\perp}^{\rm pol} \right|^2
         + \left| V_{\perp}^{F} \right|^2
      \right)
   \right] \nonumber \\
  & + &  S_{A1}S_{B1}^* \left|N_1^{\rm OMT}\right|^2
       + S_{A2}S_{B2}^* \left|N_2^{\rm OMT}\right|^2
\label{ABSparameq}
\end{eqnarray}
where the antenna signals, the polariser and OMT noises
are assumed to be uncorrelated so that 
their crossed products are zero.

Under matching conditions, for the $\parallel$ channel of the polariser
the spectral density of the noise can be
written in terms of equivalent antenna
temperature{\footnote {here  the Boltzmann costant
$k_B$ is omitted}}
\begin{eqnarray}
\left| N_{\parallel}^{\rm pol} \right|^2
      & = &   \left(
                  \frac{1}{\left| S_{\parallel} \right|^2} - 1
          \right)T_{\rm ph}^{\rm pol} \; = \;  T_{\rm noise}^{\rm pol}
     \label{VNpoleq}
\end{eqnarray}
where $T_{\rm ph}^{\rm pol}$ is the polariser physical
temperature and $T_{\rm noise}^{\rm pol}$ is the polariser equivalent noise
temperature at its input section. For $N_{\perp}$ a similar expression holds.

Because of  the presence of the
cross--coupling terms $S_{B1}$ and $S_{A2}$, 
the  noise produced by the OMT cannot be directly derived,
as for the polarizer. One should diagonalize
the transmission operator,  but this would require the knowledge of its
scattering matrix and, moreover, the diagonalization should be
frequency independent. However, under the assumption that the
off-diagonal terms are much smaller than the diagonal ones, a
conservative estimation of the OMT noise is:
\begin{eqnarray}
\left| N_1^{\rm OMT} \right|^2
       & = &     \left(
                  \frac{1}{\left| S_{A1} \right|^2 } - 1
               \right)T_{\rm ph}^{\rm OMT} 
	       \; = \; T_{\rm noise}^{\rm OMT}
\end{eqnarray}
where $T_{\rm ph}^{\rm OMT}$ is the OMT physical temperature and
$T_{\rm noise}^{\rm OMT}$is the OMT equivalent noise temperature
at its input. For the noise $N_2^{\rm OMT}$ a similar formula 
holds, where $S_{A1}$ is substituted with $S_{B2}$.

Finally, the expression for the antenna
signal is
\begin{eqnarray}
\left| V_{\parallel}^{F} \right|^2
   & = & \eta \, T_A \nonumber \\
   & = & \eta \left[
                                \left(\frac{1}{\eta} - 1
                                \right)T_{\rm ph}^{\rm horn}
                                + T_{\rm sky} + T_{\rm atm}
                  \right]  \nonumber \\
   & = & \eta \left( T_{\rm noise}^{\rm horn} +
                                   T_{\rm sky} + T_{\rm atm}\right)
\label{VNanteq}
\end{eqnarray}
where $\eta$ is the efficiency of the feed horn,
$T_{\rm ph}^{\rm horn}$ is the feed horn physical temperature,
$T_{\rm noise}^{\rm horn}$ is the feed horn equivalent noise temperature
carried to the feed horn input section, $T_A$ is the total
antenna temperature, $T_{\rm sky}$ and $T_{\rm atm}$
are the sky and the atmosphere signals, respectively.

Now, by substituting this expressions in equation~\ref{ABSparameq}, it can be 
written
\begin{eqnarray}\nonumber
A  B^*  & = & \frac{1}{2} \left( S_{A1}  S_{B1}^* + S_{A2}
                                 S_{B2}^* \right) \nonumber\\
        &   & \left[\left|S_{\parallel} \right|^2 
	            \left({1\over \left| S_{\parallel} \right|^2}
                          - 1\right)
	            T_{\rm ph}^{\rm pol}
                 + \left| S_{\perp} \right|^2 
		    \left({1\over \left| S_{\perp} \right|^2}
                          - 1\right)
		    T_{\rm ph}^{\rm pol}
   + \left(\left| S_{\parallel} \right|^2 + \left| S_{\perp} \right|^2 \right)
          \eta\, T_A \right]
    \nonumber \\
          & + &  S_{A1}S_{B1}^* \left(\frac{1}{\left|S_{A1}\right|^2}-1\right)
             T_{\rm ph}^{\rm OMT}
        +S_{A2}S_{B2}^* \left(\frac{1}{\left|S_{B2}\right|^2}-1\right)
             T_{\rm ph}^{\rm OMT} \nonumber \\
          & + & {1\over 2} \left( S_{A1}  S_{B2}^* + S_{A2}  S_{B1}^* \right)
                \left(\left| S_{\parallel} \right|^2 - \left| S_{\perp} \right|^2 \right)
                \left(\eta\, T_A - T_{\rm ph}^{\rm pol}\right)
\label{ABTeq}
\end{eqnarray}

In order to evaluate the level of the correlation product $AB^*$ it can be
assumed that
$S_{A1} \sim S_{B2}$, $S_{A2} \sim S_{B1} << 1$ and $|S_{\parallel}| \sim
|S_{\perp}|$ where they are summed. In this way,
the equation~\ref{ABTeq} becomes
\begin{eqnarray}\nonumber
A  B^*  & = & 2\, \Re(S_{A1} S_{B1}^*)
              \left[\left| S_{\parallel} \right|^2
                \left(\frac{1}{\left|S_{\parallel}\right|^2}-1\right)
               T_{\rm ph}^{\rm pol}
                    + \eta \left| S_{\parallel} \right|^2 T_A
            + \left(\frac{1}{\left|S_{A1}\right|^2}-1\right)
                 T_{\rm ph}^{\rm OMT}
            \right] +\nonumber \\
          &   &  {1\over 2} \left|S_{A1}\right|^2
                \left(\left| S_{\parallel} \right|^2 - \left| S_{\perp} \right|^2 \right)
                \left(\eta\, T_A - T_{\rm ph}^{\rm pol}\right)
\label{ABTapproxeq}
\end{eqnarray}
From equation~\ref{ABTapproxeq} one can observe that
the spurious polarisation of a real antenna system is composed
by two terms. The first one is due to the OMT, in fact this term is proportional
to the OMT cross--coupling parameter $S_{B1}$, describing the
insulation between the two OMT channels.
The second term is due to the polariser, because is proportional to
$\left| S_{\perp} \right|^2 - \left| S_{\parallel} \right|^2$.
It has to be noted that in this term $T_A$ and $T_{\rm ph}^{\rm pol}$ have
opposite sign.

Finally, this result can be carried to the feed horn input section as follows:
\begin{eqnarray}
  {A  B^*}_0  & = & {1\over\eta}
                   {1\over\left|S_{\parallel}\right|^2}
           {1\over\left|S_{A1}\right|^2}
           A  B^*,
         \label{AB0eq}
\end{eqnarray}
that through equations~\ref{VNpoleq}--\ref{VNanteq} and~\ref{ABTapproxeq}
 transforms in
\begin{eqnarray}
 {A  B^*}_0  & = & 2\,{\Re(S_{A1}S_{B1}^*)\over \left|S_{A1}\right|^2}
               \left(T_{\rm sky} + T_{\rm atm} +
                 T_{\rm noise}^{\rm Ant}
               \right) +  \nonumber \\
             &   &  {1\over 2} \left(1 - {\left|S_{\perp}\right|^2\over
                                 \left|S_{\parallel}\right|^2}\right)
                  \left(T_{\rm sky} + T_{\rm atm} +
                            T_{\rm noise}^{\rm horn} -
                            {T_{\rm ph}^{\rm pol} \over
                     \eta}
                          \right) ,
                  \label{AB0TN0eq}
\end{eqnarray}
where 
\begin{equation}
T_{\rm noise}^{\rm Ant} = T_{\rm noise}^{\rm horn} +
{1\over \eta} T_{\rm noise}^{\rm pol} + 
{1\over \eta}{1\over |S_{\parallel}|^2} T_{\rm noise}^{\rm OMT}
\end{equation}
is the noise temperature of the whole antenna system.
The equation~\ref{AB0TN0eq} provides a physical insight into the generation of
the spurious polarisation and can be directly compared with the polarised
radiation level. One can see how the various noise temperatures as well as the
sky and atmosphere emission are partially
detected as correlated signals because of the OMT 
cross--talk and the polariser
attenuation difference. In particular we can define 
two quantities that describe
the goodness of the OMT and of the polariser from the spurious polarisation
point of view:
\begin{eqnarray}
 S\!P_{\rm OMT} & = & 2\,{\Re(S_{A1}S_{B1}^*)\over \left|S_{A1}\right|^2}
                     \\
 S\!P_{\rm pol} & = & {1\over 2} \left(1 - {\left|S_{\perp}\right|^2\over
                                 \left|S_{\parallel}\right|^2}\right)
\end{eqnarray}
so that the spurious polarisation produced by the OMT and the polariser and
evaluated at the input of the antenna can be expressed as
\begin{eqnarray}
 {A  B^*}_0   =  S\!P_{\rm OMT}\left(T_{\rm sky} + T_{\rm atm} +
                 T_{\rm noise}^{\rm Ant}
               \right) +
               S\!P_{\rm pol}
                  \left(T_{\rm sky} + T_{\rm atm} +
                            T_{\rm noise}^{\rm horn} -
                            {T_{\rm ph}^{\rm pol} \over
                     \eta}
                          \right),
                  \label{AB0TNeq}
\end{eqnarray}

Moreover, following 
the {\it IAU} definition of the polarised brightness temperature
$T_b^p$ (Berkhuijsen 1975), the output for a sky signal,
in equivalent antenna temperature at the feed horn input section, is
\begin{eqnarray}
 {A  B^*}_0  & = & T_b^p (\cos(2\alpha) + j\sin(2\alpha))
\end{eqnarray}
where $\alpha$ is the polarisation angle.
Therefore, equation~\ref{AB0TNeq} provides the
estimate of the spurious polarisation level,
which can be directly compared to the polarised brightness temperature.

Since the spurious polarisation represents a polarimeter offset, the value
computed in equation~\ref{AB0TNeq} is the $T_{\rm offset}$ term of
equation~\ref{trmseq}

\subsection{Correlation of Linearly Polarised Components}

The case of a polarimeter correlating linearly polarised components ($E_X$
and $E_Y$) can be faced in a similar way, but in this case
the term due to the polariser does not exist. 
Also in this case the feed horn contribution is 
negligible, and the spurious polarisation is described by
\begin{eqnarray}
AB^* & = &  S_{A1}S_{B1}^* \left[
                               \left(\frac{1}{\left|S_{A1}\right|^2}-1\right)
                               T_{\rm ph}^{\rm OMT} + \eta\, T_A
                      \right]\nonumber\\
     & + &  S_{A2}S_{B2}^* \left[
                           \left(\frac{1}{\left|S_{B2}\right|^2}-1\right)
                               T_{\rm ph}^{\rm OMT} + \eta\, T_A
                      \right]
\label{ExEyTeq}
\end{eqnarray}
Following the assumptions adopted for the circular polarisation case,
it can be reduced to
\begin{eqnarray}
AB^* & = & 2\, \Re(S_{A1} S_{B1}^*)
                  \left[\eta\, T_A
            + \left(\frac{1}{\left|S_{A1}\right|^2}-1\right)
                 T_{\rm ph}^{\rm OMT}
          \right] .
\label{ExEyTapproxeq}
\end{eqnarray}
that at the feed horn input section yields
\begin{eqnarray}
 {AB^*}_0  & = & 2\,{\Re(S_{A1}S_{B1}^*)\over \left|S_{A1}\right|^2}
               \left(T_{\rm sky} + T_{\rm atm} +
                 T_{\rm noise}^{\rm Ant}
               \right) \nonumber \\
	      & = & S\!P_{\rm OMT} 
	       \left(T_{\rm sky} + T_{\rm atm} +
                 T_{\rm noise}^{\rm Ant}
               \right)
\label{ExEy0TNeq}
\end{eqnarray}
Obviously, here the spurious polarisation is due to the 
OMT cross--talk only.

\section{Sensitivity Degradation}\label{degrad}

The radiometer sensitivity equation (eq.~\ref{trmseq})
contains the three terms contributing to
the rms error in radio measurements. It is clear that 
both the
gain and the offset fluctuation may contribute significantly to the
sensitivity degradation,
so it is important to estimate the time scale 
on which their effects are greater than the white noise.

\subsection{Gain Fluctuations}

The noise density power spectrum of the radiometer
output (Wollack \& Pospieszalski 1998)
provides the way to compare these effects:
\begin{eqnarray}
    P(f) & = & 2{k^2\,T_{\rm sys}^2 \over \Delta \nu} +
                 T_{\rm offset}^2 \, \delta g^2(f) +
                \delta T_{\rm offset}^2(f) \nonumber\\
         & = & P_{\rm wn}(f) \;\, + P_{\rm g}(f) 
	                     \;\;\;\;\;\;\;\;\;\; + P_{\rm offset}(f)
       \label{pseq}
\end{eqnarray}
being $\delta g^2(f)$ and
$\delta T_{\rm offset}^2(f)$
the power spectra of gain and offset fluctuations, respectively.

The first term is the white noise (WN): its power spectrum is constant and
it  depends only on the istantaneous sensitivity $\sigma_0$
\begin{equation}
  P_{\rm wn}(f) = 2\,\sigma_0^2 = 2\, {k^2 T_{\rm sys}^2 \over \Delta\nu}.
\end{equation}
The second term represents the contribution of (amplifier chain)
gain fluctuations to the noise power spectrum and it depends on 
$T_{\rm offset}$. This makes a correlator more stable than
a direct detection receiver, for which $T_{\rm sys}$ must be considered
instead of $T_{\rm offset}$. Then, in a correlator receiver the contribution
of gain fluctuations is a factor $(T_{\rm offset}/T_{\rm sys})^2$ lower
than in a direct detection receiver.

Since $\delta g^2\propto f^{-\beta}$
(Wollack, 1995; Wollack \& Pospieszalski, 1998)
the second term of equation~\ref{pseq} can be written as
\begin{equation}
  P_{\rm g}(f) = 2\,\sigma_0^2\left({f_k\over f}\right)^{\beta}
  \label{pgeq}
\end{equation}
where the knee frequency $f_k$ is the frequency at which the gain
fluctuation noise equals the white noise ($\beta$ tipically is
$\sim 1$), that is
the knee frequency defines the time scale
at which gain fluctuations become higher than the instrument white noise.
Since a correlation receiver is more stable by a factor
$(T_{\rm offset}/T_{\rm sys})^2$, the relationship between
the knee frequencies of a direct detection ($f_k$) and a
correlation ($f_k^c$) receiver based on the same amplifier chain is
\begin{equation}
  f_k^c = \left(T_{\rm offset} \over T_{\rm sys}\right)^{2/\beta}
  \!\!\!\!\!f_k \;\;\simeq\;\; \left(T_{\rm offset} 
                                     \over T_{\rm sys}
			       \right)^{2} f_k
   \label{fk2fkceq}
\end{equation}
Note that $f_k^c$ scales as $\sim (T_{\rm offset} / T_{\rm sys})^2$ 
so that, for instance, a ratio  $T_{\rm offset} / T_{\rm sys} = 10^{-2}$
makes the correlator stable on a
time scale $10^{4}$ longer than an equivalent direct detection receiver.

Table~\ref{offTab} shows offset estimates, computed through
equations~\ref{AB0TNeq} and~\ref{ExEy0TNeq}, for 5 types of experiment.
L\&R@30 and L\&R@90 are ground based polarimeters
correlating the left--hand and right--hand circular
components at 30 and 90~GHz, respectively, whereas X\&Y@30 and X\&Y@90
are the cases for linear components.
For all of them attenuations have been assumed to be 
$1/\eta = 0.05$~dB,
$1/|S_C|^2 = 0.2$~dB, $1/|S_{A1}|^2 = 0.2$~dB for feed horn,
polariser and OMT, respectively, together with a common physical temperature of 20K.
Moreover, as typical values for CMB experiments are assumed: 
OMT insulation $|S_{B1}|^2 \sim -30$~dB and
polariser spurious term  $|S_{L}|^2 - |S_{C}|^2 \sim -30$~dB.
Finally, it has been considered also the SPOrt case, the only scheduled
space experiment designed to measure the microwave sky polarisation, where
physical temperatures are: $T_{\rm ph}^{\rm horn} \sim 300{\rm K}$ and 
$T_{\rm ph}^{\rm pol} = T_{\rm ph}^{\rm OMT} \sim 80{\rm K}$; 
attenuations has been assumed as for other experiments. However, since 
high care has been taken developing
the OMT, its insulation term is
$|S_{B1}|^2 \leq -60$~dB; again,
conservatively, $|S_{L}|^2 - |S_{C}|^2 \sim -30$~dB is assumed.
Also without atmospheric
contributions, the OMT is again the main offset source through the
cross--correlation of the antenna noise. 
\begin{table}
 \centering
  \caption{Offset estimates for 4 typical ground experiments at 30 and 90 GHz
           (see text) and
           for the SPOrt experiment. Different offset contributions due to the
	   OMT (upper part) and to the polariser (bottom part)
	   are expressed in mK. Atmospheric contributions are calculated for
	   typical values at Antarctica sites:
	   $T_{\rm atm}(30 {\rm
	   GHz})\sim 7 {\rm K}$ and $T_{\rm atm}(90 {\rm
	   GHz})\sim 10 {\rm K}$ (Bersanelli et al. 1995, Danese \& Partridge
	   1989).}
  \begin{tabular}{@{}lrrrrr@{}}
     \hline
            & L\&R@30 & X\&Y@30 & L\&R@90 & X\&Y@90 & SPOrt\\
     \hline
     $S\!P_{\rm OMT}T_{\rm noise}^{\rm horn}$ & $15$  & $15$  & $15$  & $15$  & $7$ \\
     $S\!P_{\rm OMT}T_{\rm noise}^{\rm pol}$  & $60$  &   -   & $60$  &   -   & $8$  \\
     $S\!P_{\rm OMT}T_{\rm noise}^{\rm OMT}$  & $63$  & $60$  & $63$  & $60$  & $8$  \\
     $S\!P_{\rm OMT}T_{\rm sky}^{\rm}$        & $170$ & $170$ & $170$ & $170$ & $5$  \\
     $S\!P_{\rm OMT}T_{\rm atm}^{\rm}$        & $442$ & $442$ & $632$ & $632$ & $0$   \\
     \hline
      Total OMT                            & $750$ & $687$ & $940$ & $877$ & $28$ \\
     \hline
     $S\!P_{\rm pol}T_{\rm noise}^{\rm horn}$       & $\sim 0$ &  - & $\sim 0$ & -  & $\sim 2$ \\
     $S\!P_{\rm pol}T_{\rm ph}^{\rm pol} / \eta$ & $10$   &  - & $10$   &  - & $39$  \\
     $S\!P_{\rm pol}T_{\rm atm}^{\rm}$              & $3.5$  &  - & $5.0$  &  - &       \\
     $S\!P_{\rm pol}T_{\rm sky}^{\rm}$              & $\sim 1$  &  - & $\sim 1$  &  - & $\sim 1$ \\
     \hline
      Total Polariser                      & $14.5$ &  -    & $16$ &    -  & $42$  \\
     \hline
     \hline
      Total (OMT+Polariser)          & $764$   & $687$ & $956$   & $877$ & $70$ \\
     \hline
\end{tabular}
\label{offTab}
\end{table}

Table~\ref{offTab} data suggest that
ground based experiments have offsets mainly
due to atmosphere emission rather than to the sky signal. Moreover, since
this offset is due to the cross--correlation of the OMT, correlating
either the circularly or the linearly polarised component is quite 
equivalent from the offset generation point of view.

\begin{table}
 \centering
  \caption{Knee frequencies for the 5 experiments of Table~\ref{offTab} 
  calculated 
  through equation~\ref{fk2fkceq}.
  The maximum allowed modulation time $t_m$ is also reported.
  $T_{\rm sys}$ and $f_k$ evaluations for ground experiments
  are based on the best
  available HEMT (NRAO) and were taken from Keating et al. 1998, Wollack \&
  Pospieszalki 1998. The SPOrt parameters have been
  calculated for TRW--MMIC amplifiers (Gaier et al. 1996; see also Cortiglioni
  et al. 1999).
           }
  \begin{tabular}{@{}lccccc@{}}
     \hline 
             & L\&R@30 & X\&Y@30 & L\&R@90 & X\&Y@90 & SPOrt\\
     \hline
     $T_{\rm sys}$ (K)      & $30$   & $30$  & $80$   &  $80$    & $150$ \\
     $T_{\rm offset}$ (mK)  & $764$  & $687$ & $956$  &  $877$   & $70$  \\
     $f_k$ (Hz)             & $50$   & $50$  & $2\times 10^3$    & $2\times 10^3$  & $50$  \\
     $f_k^c$ (Hz)           & $3.2\times 10^{-2}$ & $2.6\times 10^{-2}$ & 
                              $2.9\times 10^{-1}$ & $2.4\times 10^{-1}$ &
			      $1.1\times 10^{-5}$  \\
     $t_m$ (s)              & $31$ & $38$ & $3.5$ & $4.2$ & $9.2\times 10^{4}$   \\
     \hline
\end{tabular}
\label{kneeTab}
\end{table}

Table~\ref{kneeTab} reports the knee frequencies for the five experiments
taken as example, computed by using the offset values of Table~\ref{offTab}.
The knee frequency allows the estimate of the time 
\begin{equation}
  t_m \sim {1\over f_k^c}
\end{equation}
after that long term drifts degradate the radiometer sensitivity.
Signal modulation techniques help to overcome this degradation,
but the modulation frequency must be greater than $f_k^c$.
Table~\ref{kneeTab} shows, for example, that ground based experiments
would need modulation time $\leq 40$~s (30~GHz experiments) or 
$\leq 4$~s (90~GHz experiments). SPOrt looks much more stable and its $t_m$
parameters exceeds largely 
its modulation time (5400s i.e. its orbital period,
see Cortiglioni et al. 1999). It should be noted that such a modulation must be
inserted before the components responsible of the offset generation, to be
effective.

\subsection {Offset Fluctuations}\label{offsec}

The offset fluctuation is the second sensitivity degradation term of
equation~\ref{trmseq}. Equations~\ref{AB0TNeq} and~\ref{ExEy0TNeq}
show that it depends on both the physical temperature and
atmospheric emission fluctuations (only for ground based experiment).

Monitoring of physical temperatures and
cross--correlations with the data may allow the reduction of 
this contribution during off-line analysis, but
atmospheric fluctuations cannot be recovered and their power spectrum
adds to the white noise one. In principle,
the atmospheric emission is
unpolarised, but the spurious polarisation generated by OMT losses
fluctuates as the atmospheric emission does.
As in the case of gain fluctuations, the impact on the sensitivity can be
estimated in terms of noise power spectrum.

Atmospheric offset fluctuations can be written, following
equation~\ref{AB0TNeq}, as
\begin{eqnarray}
   \Delta T_{\rm offset}  & = &  (S\!P_{\rm OMT} + S\!P_{\rm pol})\,
                                 \Delta T_{\rm atm}\nonumber \\
                          & = &   S\, \Delta T_{\rm atm}
\end{eqnarray}
 and their contribution to the noise power spectrum is given by
\begin{eqnarray}
   P_{\rm offset}  & = & S^2 P_{\rm atm}
   \label{poffatmeq}
\end{eqnarray}
which adds to the white noise and gain fluctuations power spectra, being them
uncorrelated.

\begin{figure}
\includegraphics[scale=0.6]{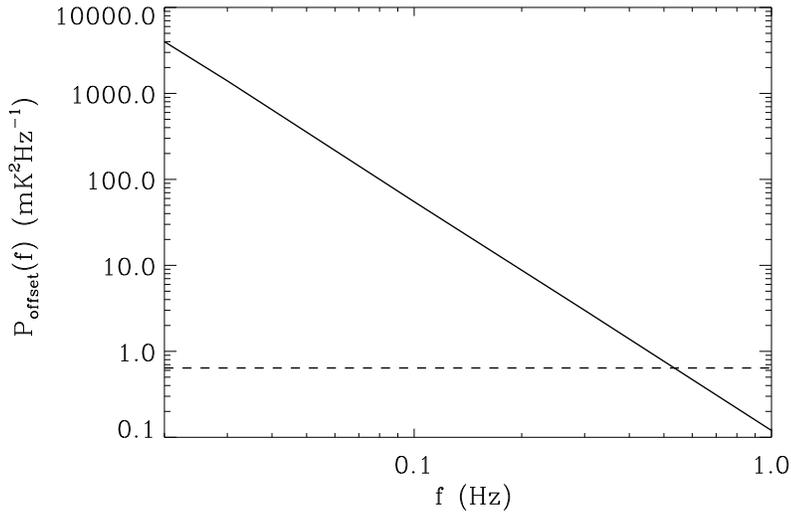}
  \caption{Estimate of the power spectrum of the offset fluctuations induced by
  atmospheric fluctuations for a 90~GHz experiment as from Table~\ref{atmTab}
  (solid line). The white noise power spectrum of a 
  typical 90~GHz amplifier is also shown
  (dashed, Wollack \& Pospieszalski 1998). The two spectra are comparable at
  $f\sim0.5$~Hz suggesting a modulation time of $t_m \sim 2$~s.
  }
\label{atmFig}
\end{figure}

Measurements taken at sea level site (Smoot et al., 1987) show that at 90~GHz
the atmospheric fluctuations follow a power law
\begin{equation}
  P_{\rm atm}(f) = \propto \, f^{-\alpha}\:\:\:\: {\rm for}\;
                             f \in [10^{-3}, 10^{-1}[
  \label{patmeq}
\end{equation}
with $\alpha = 8/3$ and normalisation 
$P_{\rm atm}(0.02{\rm Hz}) \sim 1{\rm K^2\,Hz^{-1}}$. Under these assumptions,
\begin{table}
 \centering
  \caption{Power spectrum values of both the 90~GHz atmospheric emission 
  and offset fluctuations at several frequencies $f$ obtained from
   equations~\ref{poffatmeq} and~\ref{patmeq}, assuming the normalisation 
  $P_{\rm atm}(0.02{\rm Hz}) \sim 1{\rm K^2\,Hz^{-1}}$. 
  The coefficient $S$ has been
  computed from antenna parameters as in Table~\ref{offTab}.  
           }
  \begin{tabular}{@{}ccc@{}}
     \hline 
     $f$ (Hz) & $P_{\rm atm}$~(mK$^2$Hz$^{-1}$) & $P_{\rm offset}$~(mK$^2$Hz$^{-1}$)\\
     \hline
     $0.02$   & $1.0\times 10^{6}$  & $4.3\times 10^{3}$ \\
     $0.03$   & $3.4\times 10^{5}$  & $1.4\times 10^{3}$ \\
     $0.1$   & $1.4\times 10^{4}$  & $5.5\times 10^{1}$ \\
     $0.3$   & $7.3\times 10^{2}$  & $3.0$ \\
     $1$   & $2.9\times 10^{1}$  & $0.12$ \\
     \hline
\end{tabular}
\label{atmTab}
\end{table}
Table~\ref{atmTab} shows
typical power spectrum values for both the atmosphere emission and the
corresponding polarimeter offset at different frequencies.
Similarly to the gain,
offset fluctuations begin to degradate the ideal (white noise) sensitivity
when $P_{\rm offset} > P_{\rm wn}$. Since 90~GHz radiometer
amplifiers (HEMT) have typically $P_{\rm wn} = 0.64$~mK$^2$Hz$^{-1}$
(see Wollack \& Pospieszalski
1998), the offset fluctuations get over the white noise level at
$f < 0.5$~Hz (see Figure~\ref{atmFig}). Consequently,
modulation techniques with modulation time $t_m < 2$~s  are required for longer
integration time.

Such a low $t_m$ value suggests to estimate the parameter $S$ and the insulation
of the OMT needed to ehhance it, for example, up to $t_m = 1000$~s or,
equivalently, to have $P_{\rm offset}(f=$~$10^{-3}{\rm Hz}) = P_{\rm wn}$ 
($f=10^{-3}$~Hz represents the limit of equation~\ref{patmeq}).
From equation~\ref{patmeq} and
Table~\ref{atmTab} it can be estimated 
$P_{\rm atm}(10^{-3}) \sim 2.9$~mK$^2$Hz$^{-1}$,
implying $S^2 \sim 2.2\times 10^{-10}$, that is an OMT
insulation $|S_{B1}|^2 < -100$~dB~(!) and a difference between the two
attenuations of the polariser $(|S_{\parallel}|^2 - |S_{\perp}|^2) 
< -45$~dB.

At 30~GHz the water vapour emission and its fluctuations (the main responsible
for atmospheric fluctuations) are 6 times lower than at 90~GHz. The
corresponding $P_{\rm offset}$ is 36 times lower, bringing to a slightly
better situation. In fact, using Table~\ref{atmTab} a suitable 
modulation time $t_m < 8$~s can be estimated.  

In principle, the best sites (like Antarctica), where expected
atmospheric fluctuations are lower, allow higher $t_m$, that
become comparable with values called by gain
fluctuations (see Table~\ref{kneeTab}).

\section{Conclusions}\label{conc}

The main aim of this paper is to analyse carefully the problem
related to the instrumental (spurious) polarisation, that is 
generated inside the 
instrument itself, because it can represent the major source of 
sensitivity degradation.
The reason for this is that also unpolarised signal components, 
genuinely uncorrelated, would become correlated noise 
due to losses in real receiver components. Such a correlated noise 
contributes to 
produce an offset which is responsible of sensitivity degradation (through
gain and offset fluctuations) that, 
in some cases, can be an insurmountable obstacle to the measure. That is 
the case, in particular, of the atmosphere emission, which is in principle 
unpolarised being of thermal origin. 

It has been 
recognised that CMBP investigations require state of the art receivers, 
able to give system noise, and consequently instantaneous sensitivity, 
as low as possible. However, it is well known that in any case CMBP 
measurements should require long integration times because of the 
extremely low expected signal (10\% of the anisotropy or less). As a 
consequence, the most challenging aspect of such a measurements is the 
long term stability, which is the base for useful long integration times. 

This paper has given some algorithms for spurious polarisation evaluations 
in different instrumental configurations, identifying the OMT cross--coupling
and the polariser attenuation difference as the major 
causes of correlated noise generation.
It has been also demonstrated that the antenna system is responsible of most of
the instrumental offset production, which represents the true limit to long
integration times needed by CMBP measurements.

The most relevant conclusion is that 
the best ground based experiment at 30~GHz (90~GHz) cannot 
ensure integration times longer than 40~s (4~s) without modulation. 
Alternatively, most critical components like OMT and polariser 
should have characteristics that are not
present at all in commercially available devices. 
As a consequence a significant step towards detection of
CMBP with microwave radiometers should pass through new instrumental
configurations based on custom components, 
rather than on low system noise only.
An example of such a philosophy is the SPOrt experiment, that shows how much the
long term stability can be improved by using custom designed components together
with correlation techniques.

\section*{Acknowledgments}
We thank Alessandro Orfei and the SPOrt collaboration for useful
discussions; we thank also Piermario Besso and Roberto Vallauri (CSELT)
for providing us data of the SPOrt 22~GHz horn.
This work has been partially supported by ASI contracts CNR/ASI ARS 99-15 and
CNR/ASI ARS 99-06.

\end{document}